\documentclass[12pt]{article}
\usepackage{amssymb,amsbsy,amsmath}

\newcommand{\Real}{\mathop{\mathbb R}\nolimits}           
\newcommand{\Com}{\mathop{\mathbb C}\nolimits}            
\newcommand{\ie}{\emph{i.e.}}                             
\newcommand{\eg}{\emph{e.g.}}                             
\newcommand{\cf}{\emph{cf.}}                              
\newcommand{\etc}{\emph{etc.}}                            
\newcommand{\qed}
           {\mbox{\quad\rule[-1.5pt]{.4em}{1.5ex}}}       
\newcommand{\rhs}{\emph{r.h.s.}}                          
\newcommand{\supp}{\mathop{\mathrm{supp}}\nolimits}       
\newcommand{\re}{\mathop{\mathrm{Re}}\nolimits}           

\newcommand{\sii}{\boldsymbol{L}^{2}}                     
\newcommand{\s}{\boldsymbol{L}}                           
\newcommand{\sobi}{\boldsymbol{W}_{\!0}^{1,2}}            
\newcommand{\Comp}{\boldsymbol{C}_0^\infty}               
\newcommand{\Diff}{\boldsymbol{C}^\infty}                 

\newcommand{\PF}{\textsc{Proof:\quad}}                    
\newtheorem{claim}{Claim}[section]
\newtheorem{prop}[claim]{Proposition}                     
\newtheorem{thm}[claim]{Theorem}                          
\newtheorem{corol}[claim]{Corollary}                      
\newtheorem{lemma}[claim]{Lemma}                          

\newcommand{\beqn}{\begin{eqnarray}}                      
\newcommand{\eeqn}{\end{eqnarray}}                        

\def\OMIT#1{}                                             

\begin{document}

\title{\textbf{Bound states in weakly deformed strips and layers}}
\author{D.~Borisov, P.~Exner, R.~Gadyl'shin, and D.~Krej\v{c}i\v{r}\'{\i}k}
\date{}
\maketitle

\begin{quote}
{\small {\bf Abstract.} We consider Dirichlet Laplacians on
straight strips in $\Real^2$ or layers in $\Real^3$ with a weak
local deformation. First we generalize a result of Bulla et al. to
the three-dimensional situation showing that weakly coupled bound
states exist if the volume change induced by the deformation is
positive; we also derive the leading order of the weak-coupling
asymptotics. With the knowledge of the eigenvalue analytic
properties, we demonstrate then an alternative method which makes
it possible to evaluate the next term in the asymptotic expansion
for both the strips and layers. It gives, in particular, a
criterion for the bound-state existence in the critical case when
the added volume is zero.}
\end{quote}


\setcounter{equation}{0}
\section{Introduction}

Spectra of Dirichlet Laplacians in infinitely stretched regions
such as a planar strip or a layer of a fixed width have attracted
a lot of attention recently. Of course, the problem is trivial as
long as the strip or layer is straight because then one can employ
separation of variables. However, already a local perturbation
such as bending, deformation, or a change of boundary conditions
can produce a non-empty discrete spectrum.

This effect was studied intensively in the last decade, first
because it had applications in condensed matter physics, and also
because it was itself an interesting mathematical problem. A
particular aspect we will be concerned with here is the behaviour
in the weak-coupling regime, i.e. the situation when the
perturbation is gentle.

Recall that the answer to this question depends on the type of the
perturbation. For bend strips, e.g., one can perform the Birman-Schwinger
analysis which yields the first term in the asymptotic expansion for the
gap between the eigenvalue and the threshold of the essential spectrum
\cite{DE}. It is proportional to the fourth power of the bending angle
and always positive, since any nontrivial (local) bending induces a
non-empty discrete spectrum. A local switch of the boundary condition
from Dirichlet to Neumann has a similar effect. Here the weak-coupling
behaviour was determine variationally to be governed by the fourth power
of the ``window width'' \cite{EV1} and the exact asymptotics was derived
formally in \cite{Po} by a direct application of the technique developed
in \cite{Il, Ga}. Notice that this asymptotics differs substantially from
that corresponding to a local change in the {\em mixed} boundary
conditions, where the Birman-Schwinger technique is applicable and the
leading term is a multiple of the {\em square} of the said parameter
\cite{EK}. Recall also that analogous results can be derived for layers
with locally perturbed boundary conditions where, however, the
asymptotics is exponential rather that powerlike \cite{EV2}.

The present paper deals with the case of a local {\em deformation}
of the strip or layer, which is more subtle than the bending or
boundary-condition modification. The main difference that the
effective interaction induced by a deformation can be of different
signs, both attractive and repulsive. It is easy to see by
bracketing that a bulge on a strip or layer does create bound
states while a squeeze does not. The answer is less clear for more
complicated deformations where the width change does not have a
definite sign.

The first rigorous treatment of this problem was presented in the
work of Bulla~\emph{et al}~\cite{BGRS} dealing with a local
one-sided deformation (characterized by a function $\lambda v$) of
a straight strip of a constant width~$d$. The authors found that
the added volume was decisive: a bound state exists for small
positive $\lambda$ if the area change $\lambda d\langle v \rangle$
is positive, and in that case the ground-state eigenvalue has the
following weak-coupling expansion,
\begin{equation}\label{strip-expansion}
  E(\lambda)=\kappa_1^2-\lambda^2\kappa_1^4
  \langle v \rangle^2+\mathcal{O}(\lambda^3) \,,
\end{equation}
where $\kappa_1=\frac{\pi}{d}$ is the square root of the first
transverse eigenvalue.\footnote{In fact, they assumed~$d=1$, but
it is easy to restore the strip width in their expression
obtaining eq. ~(\ref{strip-expansion}).}
On the other hand, the discrete spectrum is empty if ~$\langle v
\rangle<0$. A problem arises in the critical case, $\langle v
\rangle=0$, when the areas of the outward and inward deformation
coincide. The authors of \cite{BGRS} suggested that the analogy
with one-dimensional Schr\"odinger operators by which bound states
should exist again may be misleading due to the presence of the
higher transverse modes.

This suspicion was confirmed in \cite{EV3} where it was shown that
this is true only if the deformation was ``smeared'' enough. More
specifically, the discrete spectrum is empty if
\begin{equation} \label{crit-nonexist}
d > {4\over \sqrt 3}\:b
\end{equation}
provided $\supp v\subset [-b,b]$. On the other hand, a weakly
bound state exists if
\begin{equation} \label{crit-exist}
{\|v'\|^2\over \|v\|^2} < {6 \kappa_1^2 \over
9+\sqrt{90+12\pi^2}}\,,
\end{equation}
and in that case there are positive $c_1,\,c_2$ such that
   \begin{equation} \label{crit-asympt}
-c_1\lambda^4  \le E(\lambda)-\kappa_1^2 \le -c_2\lambda^4\,.
   \end{equation}
These results have been obtained by a variational method and they
are certainly not optimal, because there are deformed strips which
fulfill neither of the conditions (\ref{crit-nonexist}),
(\ref{crit-exist}).

A way to improve the above conclusions would be to compute the
Birman-Schwinger expansion employed in \cite{BGRS} to the second
order which becomes the leading one when the term linear in
$\lambda^2$ in~(\ref{strip-expansion}) is absent, and the
asymptotics is governed by $\lambda^4$ in correspondence
with~(\ref{crit-asympt}). This is not easy, however. The standard
technique in these situations is to map the strip in question onto
a straight one by means of suitable curvilinear coordinates. In
distinction to the bent-strip case~\cite{DE} these coordinates
typically are not locally orthogonal. Hence the transformed
Laplacian contains numerous terms which make the computation
extremely cumbersome.

After this introduction, let us describe the aim and the scope of
the present paper. The aim is twofold. First we are going to
consider an extension of the result of~\cite{BGRS} to the case of
a locally deformed layer. The result is summarized in
Theorem~\ref{thm.expansion}. In particular, we derive a
weak-coupling expansion of the ground-state eigenvalue,
\begin{equation}\label{layer-expansion}
  E(\lambda)=\kappa_1^2
  -\exp\left[2\left(-\lambda\frac{\kappa_1^2}{\pi}\langle v \rangle
  +\mathcal{O}(\lambda^2)\right)^{-1}\right]
\end{equation}
and show the analytical properties of the round-bracket expression
w.r.t.~$\lambda$. This is done in Sec.~2; the results again say
nothing about the behaviour in the critical case.

Instead of attempting to proceed further by the Birman-Schwinger
method, we demonstrate in Sec.~3 a different approach to the
weak-coupling problem. It is based on constructing the asymptotics
of a particular boundary value problem, and requires as a
prerequisite the analyticity of the function $E(\cdot)$ itself in
dimension two, and of its above mentioned constituent in dimension
three. In the present case, however, these properties are
guaranteed by \cite{BGRS} and the results of Sec.~2. The methods
allows us to recover the expansions~(\ref{strip-expansion})
and~(\ref{layer-expansion}) in a different way. What is more, we
are also able to compute higher terms, in principle of any order.
We perform the explicit computation for the second-order terms
which play role in the critical case. In particular, we made in
this way more precise the result expressed by
(\ref{crit-nonexist}) and (\ref{crit-exist}) about the critical
bound-state existence for smeared perturbations, and derive its
analog in the deformed-layer case.

\setcounter{equation}{0}
\section{Locally deformed layers}

\subsection{The curvilinear coordinates}

Let~$x=(x^1,x^2)\in\Real^2$
and~$(x,u)\in\Omega_0:=\Real^2\times(0,d)$ with~$d>0$. Given a
function~$v\in\Comp(\Real^2)$ we define the mapping ($\lambda>0$)
\begin{equation}\label{layer}
  \phi: \Omega_0\to\Real^3:
  \left\{(x,u)\mapsto\phi(x,u):=
  \left(x^1,x^2,\left(1+\lambda v(x)\right)u\right)\right\}
\end{equation}
for $\lambda>0$, which defines our deformed
layer~$\Omega_\lambda:=\phi(\Omega_0)$.

To make use of the curvilinear coordinates defined by
the mapping~$\phi$ we need the metric tensor~$G_{ij}:=\phi_{,i}.\phi_{,j}$
of the deformed layer. It can be seen easily to be of the form
\begin{equation}\label{metric}
  (G_{ij})=
  \left(
  \begin{array}{ccc}
    1+\lambda^2 v_{,1}^2 u^2
    & \lambda^2 v_{,1} v_{,2} u^2
    & \lambda v_{,1} (1+\lambda v) u \\
    \lambda^2 v_{,1} v_{,2} u^2
    & 1+\lambda^2 v_{,2}^2 u^2
    & \lambda v_{,2} (1+\lambda v) u \\
    \lambda v_{,1} (1+\lambda v) u
    & \lambda v_{,2} (1+\lambda v) u
    & (1+\lambda v)^2
  \end{array}
  \right)\,,
\end{equation}
where $v_{,\mu}$ means the derivative w.r.t. $x^\mu$,
and its determinant is $G:=\det(G_{ij})=(1+\lambda v)^2$.

In view of the inverse function theorem, the mapping~$\phi$
defining the layer will be diffeomorphism provided~$\lambda
\|v_-\|_\infty<1$, where we put conventionally
$v_-:=\max\{0,-v\}$. For a sign-changing $v$, this is a nontrivial
restriction which is satisfied, however, when $\lambda$ is small
enough. That is just the case we are interested in.

We will also need the contravariant metric tensor, in other words
the inverse matrix
\begin{equation}\label{inversemetric}
  (G^{ij})=
  \left(
  \begin{array}{ccc}
    1
    & 0
    & -\frac{\lambda v_{,1} u}{1+\lambda v} \\
    && \\
    0
    & 1
    & -\frac{\lambda v_{,2} u}{1+\lambda v} \\
    && \\
    -\frac{\lambda v_{,1} u}{1+\lambda v}
    & -\frac{\lambda v_{,2} u}{1+\lambda v}
    & \frac{1+\lambda^2 |\nabla v|^2 u^2}{(1+\lambda v)^2}
  \end{array}
  \right)
\end{equation}
and the following contraction identities
\begin{equation}\label{contraction}
  G^{\mu j}_{\phantom{\mu},j}=-\frac{\lambda v_{,\mu}}{1+\lambda v},
  \qquad
  G^{3j}_{\phantom{3},j}=-\frac{\lambda \Delta v \,u}{1+\lambda v}
  +\frac{3 \lambda^2 |\nabla v|^2 u}{(1+\lambda v)^2}\,,
\end{equation}
where conventionally summation is performed over repeated indices,
and we denote~$|\nabla v|^2:=v_{,1}^2+v_{,2}^2$ and $\Delta
v:=v_{,11}+v_{,22}$. Another convention concerns the range of the
indices, which is ~$1,2$ for Greek and~$1,2,3$ for Latin indices.
The indices are at that associated with the above coordinates by
\mbox{$(1,2,3)\leftrightarrow (x^1,x^2,u)$}.

\subsection{The straightening transformation}

As mentioned in the introduction the main object of our study is
the Dirichlet Laplacian ~$-\Delta_D^{\Omega_\lambda}$
on~$\sii(\Omega_\lambda)$. If we think of a quantum particle
living in the region $\Omega_\lambda$ with hard walls and exposed
to no other interaction, $-\Delta_D^{\Omega_\lambda}$ will be its
Hamiltonian up to a multiplicative constant; we can get rid of the
latter by setting the Planck's constant $\hbar=1$ and the
effective mass $m_*=\frac{1}{2}$. Mathematically speaking,
$-\Delta_D^{\Omega_\lambda}$ is defined for an open
set~$\Omega_\lambda\subset\Real^3$ as the Friedrichs extension of
the free Laplacian with the domain~$\Comp(\Omega)$ --
\cf~\cite[Sec.~XIII.15]{RS4}. Moreover, since the smooth boundary
of $\Omega_\lambda$ has the segment property,
$-\Delta_D^{\Omega_\lambda}$ acts simply as $\psi \mapsto
-\psi_{,jj}$ with the Dirichlet b.c. at $\partial\Omega_\lambda$.

A natural way to investigate the Hamiltonian is to introduce
the unitary transformation
$
  U:\sii(\Omega_\lambda)\to\sii(\Omega_0):
  \{\psi \mapsto U\psi:=G^\frac{1}{4}\psi\circ\phi\}
$
and to investigate the unitarily equivalent operator
\begin{equation}\label{Hamiltonian}
  H_\lambda:=U(-\Delta_D^{\Omega_\lambda})U^{-1}
  =-G^{-\frac{1}{4}}\partial_i G^{\frac{1}{2}}
  G^{ij}\partial_j G^{-\frac{1}{4}}
\end{equation}
with the form domain~$Q(H_\lambda)=\sobi(\Omega_0)$ instead of
$-\Delta_D^{\Omega_\lambda}$. As usual in such situations, the
``straightened'' region is geometrically simpler and the price we
pay is a more complicated form of the operator
(\ref{Hamiltonian}).

To make it more explicit, put~$F:=\ln G^\frac{1}{4}$. Commuting
$G^{-\frac{1}{4}}$ with the gradient components, we cast the
operator~(\ref{Hamiltonian}) into a form which has a simpler
kinetic part,
$$
  H_\lambda=-\partial_i G^{ij}\partial_j+V
  =-G^{ij}\partial_i\partial_j-G^{ij}_{\phantom{i},j}\partial_i+V\,,
$$
but contains an effective potential,
$$
  V:=(G^{ij}F_{,j})_{,i}+F_{,i}G^{ij}F_{,j}
  =G^{ij}F_{,ij}+G^{ij}_{\phantom{i},j}F_{,i}+G^{ij}F_{,i}F_{,j}.
$$
If we now employ the particular form~(\ref{metric}) of the metric
tensor together with~(\ref{inversemetric}), (\ref{contraction}),
we can write
\begin{eqnarray*}
H_\lambda
&=& -\partial_1^2-\partial_2^2
  -\frac{1+\lambda^2 |\nabla v|^2 u^2}{(1+\lambda v)^2} \partial_3^2
  +\frac{2\lambda v_{,1} u}{1+\lambda v} \partial_1 \partial_3
  +\frac{2\lambda v_{,2} u}{1+\lambda v} \partial_2 \partial_3 \\
&& +\frac{\lambda v_{,1}}{1+\lambda v} \partial_1
  +\frac{\lambda v_{,2}}{1+\lambda v} \partial_2
  +\left(\frac{\lambda \Delta v \,u}{1+\lambda v}
  -\frac{3\lambda^2 |\nabla v|^2 u}{(1+\lambda v)^2}\right) \partial_3
  +V
\end{eqnarray*}
with
$$
  V=\frac{\lambda \Delta v}{2}
  -\frac{\lambda^2 v \Delta v}{2(1+\lambda v)}
  -\frac{3\lambda^2 |\nabla v|^2}{4(1+\lambda v)^2}.
$$
For our purpose it useful to rewrite this expression further in a
form sorted w.r.t. to the powers of~$\lambda$:
\begin{eqnarray*}
\lefteqn{H_\lambda=-\Delta_D^{\Omega_0} +\lambda
  \bigg[2v\partial_3^2
  +2v_{,1}u\partial_1\partial_3+2v_{,2}u\partial_2\partial_3
  +v_{,1}\partial_1+v_{,2}\partial_2 } \\
&&\phantom{-\lambda^2} +(\Delta v) \,u \partial_3
  +\frac{\Delta v}{2} \bigg] \\
&& -\lambda^2 \Bigg[\frac{3v^2+|\nabla v|^2 u^2+2\lambda v^3}
  {(1+\lambda v)^2}\, \partial_3^2
  +\frac{2v v_{,1} u}{1+\lambda v}\, \partial_1\partial_3
  +\frac{2v v_{,2} u}{1+\lambda v}\, \partial_2\partial_3 \\
&& \phantom{-\lambda^2}
  +\frac{v v_{,1}}{1+\lambda v}\, \partial_1
  +\frac{v v_{,2}}{1+\lambda v}\, \partial_2
  +\left(\frac{v (\Delta v) \,u}{1+\lambda v}
  +\frac{3|\nabla v|^2 u}{(1+\lambda v)^2}\right) \partial_3 \\
&& \phantom{-\lambda^2}
  +\frac{v \Delta v}{2(1+\lambda v)}
  +\frac{3|\nabla v|^2}{4(1+\lambda v)^2}
  \Bigg]
\end{eqnarray*}
In analogy with~\cite{BGRS}, we thus get the following formula for
the ``straightened'' operator,
\begin{equation}\label{HamAB}
  H_\lambda=H_0+\lambda \sum_{n=1}^3 A_n^* B_n
  +\lambda^2 \sum_{n=4}^7 A_n^* B_n,
\end{equation}
where each of the~$A_n$'s and~$B_n$'s is a first-order
differential operator with compactly supported coefficients and
\begin{align*}
  A_1^* & := 2v\partial_3
  & B_1 & := \omega \partial_3 \\
  A_2^* & := \Delta v
  & B_2 & := \omega \left(u\partial_3+\frac{1}{2}\right) \\
  A_3^* & := \left(2u\partial_3+1\right)\omega
  & B_3 & := v_{,1}\partial_1+v_{,2}\partial_2 \\
  A_4^* & := -\frac{3v^2+|\nabla v|^2 u^2+2\lambda v^3}{(1+\lambda v^3)}
             \,\partial_3
  & B_4 & := \omega\partial_3 \\
  A_5^* & := -\frac{v \Delta v}{1+\lambda v}
  & B_5 & := \omega \left(u\partial_3+\frac{1}{2}\right) \\
  A_6^* & := -\frac{3 |\nabla v|^2}{(1+\lambda v)^2}
  & B_6 & := \omega \left(u\partial_3+\frac{1}{4}\right) \\
  A_7^* & := -\frac{2u\partial_3+1}{1+\lambda v}\, v
  & B_7 & := v_{,1}\partial_1+v_{,2}\partial_2
\end{align*}
with~$\omega\in\Comp(\Real^2)$ such that~$\omega\equiv 1$
on~$\supp v$. We define a pair of operators
$C_\lambda,D:\sii(\Omega_0)\to\sii(\Omega_0)\otimes\Com^7$ by
$$
  \begin{array}{l}
    \varphi\mapsto (C_\lambda\varphi)_n:=
    \left\{
    \begin{array}{rl}
      A_n\varphi         & n=1,2,3 \\
      \lambda A_n\varphi & n=4,\dots,7
    \end{array}
    \right.
    \\
    \\
    \varphi\mapsto (D\varphi)_n:=B_n\varphi
    \quad n=1,\dots,7
  \end{array}
$$
then~(\ref{HamAB}) finally becomes~$H_\lambda=H_0+\lambda
C_\lambda^* D$.

\subsection{Weak coupling analysis}

First we note that since the our layer is deformed only locally,
we have
$$
  \sigma_\mathrm{ess}(-\Delta_D^{\Omega_\lambda})
  =\sigma_\mathrm{ess}(-\Delta_D^{\Omega_0})
  =[\kappa_1^2,\infty)\,.
$$
This is easy to see, for instance, by using a bracketing to show
that $\inf\sigma_\mathrm{ess}(-\Delta_D^{\Omega_\lambda})
=\kappa_1^2\;$ -- \cf~\cite{DEK} -- while the opposite inclusion
is obtained by constructing an appropriate Weyl sequence. We use
the notation ~$\kappa_j^2:=(\frac{\pi}{d}j)^2$ for the eigenvalues
of the transverse operator~$(-\partial_3^2)^D$; the corresponding
eigenfunctions are denoted by~$\chi_j$, and their explicit form is
$$
  \chi_j(u)=\sqrt{\frac{2}{d}}\,\sin\kappa_n u\,.
$$
Next we define~$K_\lambda^\alpha:=\lambda D (H_0-\alpha^2)^{-1}
C_\lambda^*$. We are interested in (positive)
eigenvalues~$E(\lambda)=:\alpha^2$ of~$H_\lambda$ below the lowest
transverse mode, hence we choose $\alpha\in[0,\kappa_1)$. Our
basic tool is the following classical result --
\cf~\cite[Lemma~2.1]{BGRS}:
\begin{prop}[Birman-Schwinger principle]
$$
  \alpha^2\in\sigma_\mathrm{disc}(H_\lambda)
  \Longleftrightarrow
  -1\in\sigma_\mathrm{disc}(K_\lambda^\alpha)
$$
\end{prop}
\PF If~$K_\lambda^\alpha\psi=-\psi$, then $\varphi:=-\lambda
(H_0-\alpha^2)^{-1} C_\lambda^*\psi$ is easily checked to satisfy
\mbox{$H_\lambda\varphi=\alpha^2\varphi$}. Conversely, if
\mbox{$H_\lambda\varphi=\alpha^2\varphi$}, we have $\varphi\in
Q(H_\lambda)\subset D(D)$, so $\psi:=D\varphi$ is in
$\sii(\Omega_0)$ and $K_\lambda^\alpha\psi=-\psi$.\qed
\vspace{2mm}

To make use of the above equivalence, we have to analyze the
structure of $K_\lambda^\alpha$. Let $R_0(\alpha)
:=(H_0-\alpha^2)^{-1}$ be the free resolvent corresponding
to~$H_0$. Using the transverse-mode decomposition and the fact
that $H_0=-\Delta^{\Real^2}\otimes I_1+I_2 \otimes
(-\partial_3^2)^D$, we can express the integral kernel of~$R_0$,
$$
  R_0(x,u,x',u';\alpha)=\sum_{j=1}^\infty
  \chi_j(u) \, r_j(x,x';\alpha) \, \chi_j(u')
$$
where $r_j(x,x';\alpha)$ is the kernel of~$(-\Delta^{\Real^2}
+\kappa_j^2-\alpha^2)^{-1}$ in~$\sii(\Real^2)$. We define
$k_j(\alpha)^2:=\kappa_j^2-\alpha^2$. The free kernel~$r_j$ can be
expressed in terms of Hankel's functions -- \cf~\cite[Chap.~I.5]{AGH} --
which are related to Macdonald's  functions by~\cite[9.6.4]{AS}, so
finally we arrive at the formula
$$
  R_0(x,u,x',u';\alpha)=\frac{1}{2\pi}\sum_{j=1}^\infty
  \chi_j(u) \,
  K_0\left(k_j(\alpha)|x-x'|\right) \,
  \chi_j(u')\,.
$$
Now we want to split the singular part of~$R_0^\alpha$; we write
$K_\lambda^\alpha=\hat{L}_\lambda+\hat{M}_\lambda$ where
$\hat{L}_\lambda:=\lambda D L_\alpha C_\lambda^*$ contains the
singularity:
$$
  L_\alpha(x,u,x',u'):=-\frac{1}{2\pi} \,
  \chi_1(u) \, \ln k_1(\alpha) \, \chi_1(u')
$$
diverges logarithmically as $\alpha\to\kappa_1-$. The regular part
$\hat{M}_\lambda=\lambda D M_\alpha C_\lambda^*$ consists of two terms,
$M_\alpha=N_\alpha+R_0^\bot(\alpha)$, where the operator~$R_0^\bot$ is
defined as the projection of the resolvent on higher transverse modes
$$
  R_0^\bot(x,u,x',u';\alpha):=\frac{1}{2\pi}\sum_{j=2}^\infty
  \chi_j(u) \,
  K_0(k_j(\alpha)|x-x'|) \,
  \chi_j(u'),
$$
and the remaining term is therefore
$$
  N_\alpha(x,u,x',u'):=\frac{1}{2\pi}
  \chi_1(u) \,
  \Big(K_0(k_1(\alpha)|x-x'|)+\ln k_1(\alpha)\Big)
  \chi_1(u').
$$
Put~$w^{-1}:=\ln k_1(\alpha)$. The next step in the BS method is to show
the boundedness and the analyticity (w.r.t.~$w$) of the regular part
of~$K_\lambda^\alpha$. A  more difficult part of this task concerns the
operator containing $N_{\alpha}$ where we have to take a different route
than that used in \cite{BGRS}.

First we note that while the Hilbert-Schmidt norm is suitable for
estimating the operator~$N_\alpha$, it fails when the latter is
sandwiched between $\lambda D$ and $C^*_\lambda$. More specifically,
using the regularity and compact support of the functions involved one
could transform $\lambda D N_\alpha C_\lambda^*$ into an integral
operator via integration by parts, but the obtained kernel has a
singularity which is not square integrable. Hence we use instead the
``continuous'' version of the {\em Schur-Holmgren bound.} Since it seems
to be less known than its discrete analogue ~\cite[Lemma~C.3]{AGH},
\cite[Thm.~7.1.9]{Maddox}, we present it here with the proof.
\begin{lemma}\label{Holmgren}
Suppose that~$M$ is an open subset of~$\Real^n$ and let
~$K:\sii(M)\to\sii(M)$ be an integral operator with the kernel
$K(\cdot,\cdot)$. Then
$$
  \|K\| \leq \|K\|_\mathrm{SH}
  :=\left(\sup_{x\in M}\int_M |K(x,x')| dx'
  \ \sup_{x'\in M} \int_M |K(x,x')| dx
  \right)^\frac{1}{2}.
$$
\end{lemma}
\PF The claim follows from the inequality
\begin{equation}\label{estimate}
  \|K\|_{p,p} \leq  \|K\|_{1,1}^{1/p} \,
  \|K\|_{\infty,\infty}^{1/q} \,,
\end{equation}
where~$K$ is now an integral operator on~$\s^p(M)$,
$\;p^{-1}+q^{-1}=1$, and
$$
  \|K\|_{\infty,\infty}:=\sup_{x\in M}
  \int_M |K(x,x')| \,dx',
  \quad
  \|K\|_{1,1}:=\sup_{x'\in M}
  \int_M |K(x,x')| \,dx.
$$
If~$K$ is bounded for~$p=1,\infty$, we can prove ~(\ref{estimate})
for the other~$p$ by an interpolation argument adapted from the
discrete case \cite{Maddox}. By H\"older's inequality
\begin{eqnarray*}
\lefteqn{\left|\int_M K(x,x')\psi(x') dx'\right| \leq
  \int_M |K(x,x')|^\frac{1}{p} |K(x,x')|^\frac{1}{q} |\psi(x')| \,dx'} \\
&& \leq \left(\int_M |K(x,x')| |\psi(x')|^p dx'\right)^\frac{1}{p}
  \int_M |K(x,x')| \,dx'\,,
\end{eqnarray*}
so we can easily estimate the~$\s^p$-norm of~$K\psi$,
\begin{eqnarray*}
  \|K\psi\|_p^p
&=& \int_M dx \left|\int_M K(x,x') \psi(x') \,dx' \right|^p \\
&\leq& \|K\|_{\infty,\infty}^{p/q}
  \int_M dx \int_M |K(x,x')| |\psi(x')|^p \,dx' \\
&\leq& \|K\|_{\infty,\infty}^{p/q}
  \int_M dx'\,|\psi(x')|^p \int_M dx\,|K(x,x')| \\
&\leq& \|K\|_{\infty,\infty}^{p/q}
  \|K\|_{1,1} \|\psi\|_p^p\,,
\end{eqnarray*}
which yields the result.\qed \vspace{2mm}

Recall that~$\|\cdot\|_\mathrm{SH}$ is \emph{not} a norm and that it
simplifies for the symmetric kernels, $\|K\|_\mathrm{SH} =\sup_{x\in
M}\int_M |K(x,x')| \,dx'$. We are now ready to prove the following key
result.
\begin{lemma}\label{analyt}
$w \mapsto \hat{M}(\alpha(w))$ is a bounded and
analytic operator-valued function, which can be continued from
$\{w\in\Com | \re w<0\}$ to a region that includes~$w=0$.
\end{lemma}
\PF As in~\cite[Lemma~2.2]{BGRS}, let~$\mathcal{H}_1\subset\sii(\Omega_0)$
be the space of~$\sii(\Omega_0)$ functions of the form~$\varphi\chi_1$,
where $\varphi\in\sii(\Real^2)$. Let further~$\mathcal{P}_1$ be the
projection onto this subspace, and~$\mathcal{P}_1^\bot:=I-\mathcal{P}_1$
the projection onto its orthogonal complement in~$\sii(\Omega_0)$.
Then~$R_0^\bot(\alpha) \equiv R_0(\alpha)\mathcal{P}_1^\bot$ has an
analytic continuation into the region $\{\alpha\in\Com |
\alpha^2\in\Com\setminus[3\kappa_1^2,\infty)\}$ since the lowest point in
the spectrum of $H_0\mathcal{P}_1^\bot \upharpoonright
\mathcal{P}_1^\bot\sii(\Omega_0)$ is~\mbox{$\kappa_2^2-\kappa_1^2$}. This
region includes the domain~$[0,\kappa_1)$ actually considered. To
accommodate the extra factors~$D,C_\lambda^*$, we introduce the quadratic
form
$$
  b_\alpha(\phi,\psi) := (\phi,D R_0^\bot(\alpha) C_\lambda^* \psi) =
  (R_0^\bot(\alpha)^\frac{1}{2} \mathcal{P}_1^\bot D^* \phi,
  R_0^\bot(\alpha)^\frac{1}{2} \mathcal{P}_1^\bot C_\lambda \psi)\,.
$$
To check boundedness of this form, it is therefore sufficient to verify
that~$R_0^\bot(\alpha)^\frac{1}{2} \mathcal{P}_1^\bot D^*$
and~$R_0^\bot(\alpha)^\frac{1}{2} \mathcal{P}_1^\bot C_\lambda^*$ are
bounded operators. We shall check it for their adjoints. To this purpose,
it is enough to show that~$C_\lambda \mathcal{P}_1^\bot$ and~$D
\mathcal{P}_1^\bot$ are $(R_0^\bot(\alpha)^{-\frac{1}{2}}
\mathcal{P}_1^\bot)$-bounded, \ie, that there exist positive~$a,b$ such
that
$$
  \forall\psi\in Q(H_\lambda): \qquad
  \|C_\lambda \mathcal{P}_1^\bot \psi\|
  \leq a \|R_0^\bot(\alpha)^{-\frac{1}{2}} \mathcal{P}_1^\bot \psi\|
  + b \|\psi\|\,,
$$
and similarly for~$D \mathcal{P}_1^\bot$. However,
\begin{eqnarray*}
  \|\nabla \mathcal{P}_1^\bot \psi\|^2
&=& \|(H_0+1)^\frac{1}{2} \mathcal{P}_1^\bot \psi\|^2
  -\|\mathcal{P}_1^\bot \psi\|^2 \\
  \|(H_0+1)^\frac{1}{2} \mathcal{P}_1^\bot \psi\|
&\leq& \|(H_0-\alpha^2)^\frac{1}{2} \mathcal{P}_1^\bot \psi\|
  +\sqrt{1+\alpha^2} \|\mathcal{P}_1^\bot \psi\| \\
&\leq& \|R_0^\bot(\alpha)^{-\frac{1}{2}} \mathcal{P}_1^\bot \psi\|
  +\sqrt{1+\alpha^2} \|\psi\|\,.
\end{eqnarray*}
Here $\nabla$ means the gradient in the variables~$(x,u)$) through which
all the actions of~$C_\lambda, D$ can be estimated, \eg,
$|(C_\lambda\psi)_1| \equiv |A_1\psi| \leq 2\|v\|_\infty |\nabla\psi|$,
\etc\/ In the same way, one verifies the analyticity of the
operator-valued function~$D R_0^\bot(\alpha) C_\lambda^*$, which is
equivalent to the analyticity of the complex-valued function
$\alpha\mapsto b_\alpha(\cdot,\cdot)$.

Consider next the regular part of~$R_0(\alpha)\mathcal{P}_1$ containing
the operator~$N_\alpha$. Let~$h$ be a~$\Diff$-function of compact support
in~$\Real^2$. As pointed out above, using integration by parts and the
explicit form of the operators~$C_\lambda, D$ one sees that it is
sufficient to check the boundedness and analyticity of~$h n_\alpha h$
and~$h n_{\alpha,\mu} h$, where
\begin{eqnarray*}
  n_\alpha(x,x')
&:=&\frac{1}{2\pi} K_0(k_1(\alpha)|x-x'|)+\ln k_1(\alpha)\,, \\
  n_{\alpha,\mu}(x,x')
&=&-\frac{1}{2\pi}\ \frac{x^\mu-{x'}^\mu}{|x-x'|}\
  k_1(\alpha) K_1(k_1(\alpha)|x-x'|)\,;
\end{eqnarray*}
recall that~$_{,\mu}$ means the derivative w.r.t.~$x^\mu$ and~$K_0'=-K_1$
holds true -- \cf~\cite[9.6.27]{AS}. We will use the following estimates
which are valid for the Macdonald functions~\cite[9.6--7]{AS} with any $
z\in (0,\infty)$:
\begin{align*}
  |(K_0(z)+\ln z)e^{-z}|
  & \leq c_1\,, &
  |K_1(z)-z^{-1}|
  & \leq c_2\,, \\
  |[K_1(z)-z (K_0(z)+K_2(z))/2]|
  & \leq c_3\,, &
  |z K_1(z)|
  & \leq 1\,.
\end{align*}
Passing to the polar coordinates,
$$
  x^\mu-{x'}^\mu=\left(\rho \cos\varphi,\rho \sin\varphi\right)\,,
  \quad \rho_m:=\sup_{x,x'\in\supp h}|x-x'|\,,
$$
we check the finiteness of the Schur-Holmgren bounds:
\begin{eqnarray*}
\|h n_\alpha h\|_\mathrm{SH}
&=& \sup_{x\in\Real^2} |h(x)|
  \int_{\Real^2} |m_1(x,x';\alpha) h(x')| \,dx' \\
&\leq&
   c_1 \|h\|_\infty^2 \int_0^{\rho_m} e^{k_1(\alpha)\rho} \rho\,d\rho
  +\int_0^{\rho_m} |\ln\rho|\,\rho\,d\rho \\
&\leq&
   c_1 \|h\|_\infty^2\,\rho_m
  \left(\rho_m e^{\kappa_1 \rho_m}+\max\{e^{-1},\rho_m \ln\rho_m\}\right) \,,\\
\|h n_{\alpha,\mu} h\|_\mathrm{SH}
&\leq& \|h\|_\infty^2 \int_0^{\rho_m} \frac{\rho\,d\rho}{\rho}
  =\|h\|_\infty^2\,\rho_m.
\end{eqnarray*}
Concerning the analyticity, one should investigate the
com\-plex-valued functions $w \mapsto\left(\phi,h n_{\alpha(w)}
h\,\psi\right)$ and $w \mapsto\left(\phi,h n_{\alpha(w),\mu}
h\,\psi\right)$, where~$\phi,\psi$ are arbitrary vectors
of~$\sii(\Omega_0)$. Using the Schwarz inequality, it is
sufficient to check the finiteness of norms of the complex
derivative w.r.t.~$w$ of the corresponding operator-valued
functions. Since~$K_1'=-(K_0+K_2)/2$ by~\cite[9.6.29]{AS}
and~$k_1(\alpha(w))=e^{w^{-1}}$, we put $z:=k_1(\alpha(w))|x-x'|$
and write
{\setlength\arraycolsep{2pt}
\begin{eqnarray*}
\frac{d n_{\alpha(w)}}{dw}(x,x')
&=& \frac{1}{2\pi}
  \frac{z}{w^2} \left(K_1(z)-\frac{1}{z}\right)\,, \\
\frac{d n_{\alpha(w),\mu}}{dw}(x,x')
&=& \frac{1}{2\pi} \frac{x^\mu-{x'}^\mu}{|x-x'|}
  \frac{e^{w^{-1}}}{w^2}
 \left[K_1(z)-\frac{z}{2}\big(K_0(z)+K_2(z)\big)
  \right]\,.
\end{eqnarray*}}

\noindent Using now the inequality~$w^{-2}e^{w^{-1}}\leq c_4$
for~$w\in(-\infty,0)$, we are able to estimate the Schur-Holmgren
bounds:
$$
  \left\|h \frac{d n_{\alpha(w)}}{dw} h\right\|_\mathrm{SH}
  \leq c_2 c_4 \|h\|_\infty^2\,\rho_m^2\,,
  \quad
  \left\|h \frac{d n_{\alpha(w),\mu}}{dw} h\right\|_\mathrm{SH}
  \leq c_3 c_4 \|h\|_\infty^2\,\rho_m^2\,.
$$
Thus the derivatives are bounded for~$w\in(-\infty,0)$, and since the
limits as~$w$ tends to zero make sense, we can continue the function
analytically to~$w=0$.\qed

Now we are in position to follow the standard Birman-Schwinger scheme to
derive the weak-coupling expansion. Eigenvalues of~$H_\lambda$ correspond
to singularities of the operator-valued function
$(I+K_\lambda^\alpha)^{-1}$ which we can express as
\begin{equation}\label{K->M,L}
  (I+K_\lambda^\alpha)^{-1}
  =\left[I+(I+\hat{M}_\lambda)^{-1}\hat{L}_\lambda\right]^{-1}
  (I+\hat{M}_\lambda)^{-1}.
\end{equation}
Owing to~Lemma~\ref{analyt}, $\|\hat{M}_\lambda\|$ is finite and we can
choose~$\lambda$ sufficiently small to have $\|\hat{M}_\lambda\|<1$; then
the second term at the \rhs\/ of~(\ref{K->M,L}) is a bounded operator. On
the other hand, $(I+\hat{M}_\lambda)^{-1}\hat{L}_\lambda$ is a rank-one
operator of the form $(\psi,\cdot)\varphi$, where
\begin{eqnarray*}
  \bar{\psi}(x,u)
& := & -\frac{\lambda}{2\pi} \, \ln k_1(\alpha) \,\chi_1(u) C_\lambda^* \,,\\
  \varphi(x,u)
& := & \left[(I+\hat{M}_\lambda)^{-1}  D\,\chi_1\right](x,u)\,,
\end{eqnarray*}
so it has just one eigenvalue which is
$$
  (\psi,\varphi)=-\frac{\lambda}{2\pi} \, \ln k_1(\alpha)
  \int_0^d \!\! \int_{\Real^2}  \chi_1(u) \,C_\lambda^*
   \left[(I+\hat{M}_\lambda)^{-1} D\,\chi_1\right](x,u) \, dx\,du\,.
$$
Putting it equal $-1$ we get an implicit equation, $F(\lambda,w)=0$, with
\begin{equation}\label{ImplicitEq}
  F(\lambda,w):=w-\frac{\lambda}{2\pi}
  \int_0^d \!\! \int_{\Real^2}  \chi_1(u) \,C_\lambda^*
  \left[(I+\hat{M}_\lambda)^{-1} D\,\chi_1\right](x,u) \, dx\,du \,,
\end{equation}
where~$\hat{M}_\lambda$ has to be understood as a function both
of~$\lambda$ and~$w$. Expanding \mbox{$(I+\hat{M}_\lambda)^{-1}$} into
the Neumann series we find
$$
  F_{,w}(0,0)=1\not=0\,,
  \qquad
  F_{,\lambda}(0,0)=-\frac{1}{2\pi} \left(\chi_1,C_0^* D\,\chi_1\right)\,,
$$
and by~Lemma~\ref{analyt} we know that~$F(\lambda,w)$ is jointly analytic
in~$\lambda,w$. In view of the implicit function theorem $w=w(\lambda)$ is
then an analytic function and we can compute the first term in its Taylor
expansion:
$$
  \frac{dw}{d\lambda}(0)=-\frac{F_{,\lambda}(0,0)}{F_{,w}(0,0)}=
  \frac{1}{2\pi} \left(\chi_1,C_0^* D\,\chi_1\right)\,.
$$
But~$(C_0)_n=0$ for~$n=4,\dots,7$,  $B_3\chi_1=0$,
and $(A_2\chi_1,B_2\chi_1)=0$
since \mbox{$\int_{\Real^2} \Delta v=0$}. It follows that
\begin{equation}\label{der}
  \frac{dw}{d\lambda}(0)
  =\frac{1}{2\pi} \left(A_1\chi_1,B_1\chi_1\right)
  =-\frac{1}{\pi} \int_0^d \chi_1'(u)^2 \,du \int_{\Real^2}\!\! v(x)\,dx
  =-\frac{\kappa_1^2}{\pi}\, \langle v \rangle,
\end{equation}
where we have employed the symbol~$\langle v \rangle:=\int_{\Real^2}
v(x)\,dx$.

We note that~$\alpha^2\to\kappa_1^2-$ holds as~\mbox{$\lambda\to 0+$},
and consequently, $k_1(\alpha)\to 0+$. Thus~$w(0)=0$ is well defined
because \mbox{$w=(\ln k_1(\alpha))^{-1}$} by definition. Furthermore, the
solution~$\alpha^2$ clearly represents an eigenvalue if and only if~$w$ is
strictly negative for~$\lambda$ small. A sufficient condition for that is
that the first term of the expansion of~$w(\lambda)$ is strictly
negative; due to~(\ref{der}) it happens if~$\langle v \rangle$ is strictly
positive. Summing up the discussion, we get the announced
three-dimensional analogue to Theorem~1.2 in~\cite{BGRS}:
\begin{thm}\label{thm.expansion}
Let ~$\Omega_\lambda$ be given by~(\ref{layer}),
where~$v\in\Comp(\Real^2)$ satisfies~$\langle v \rangle>0$. Then for all
sufficiently small positive~$\lambda$, $-\Delta_D^{\Omega_\lambda}$ has a
unique eigenvalue~$E(\lambda)$ in~$[0,\kappa_1^2)$, which is simple and
can be expressed as
$
  E(\lambda)=\kappa_1^2-e^{2 w(\lambda)^{-1}},
$
where~$\lambda\mapsto w(\lambda)$ is an analytic function. Moreover, the
following asymptotic expansion is valid:
$$
  w(\lambda)=-\lambda\,\frac{\kappa_1^2}{\pi}\,\langle v \rangle
  +\mathcal{O}(\lambda^2)\,.
$$
\end{thm}

\setcounter{equation}{0}
\section{An alternative method}

Now we will derive the weak-coupling expansion by constructing the
asymptotics for singularities in a particular boundary value problem. This
approach enables us to derive easily higher terms of the expansion. At
the same time it allows a unified treatment for different dimensions; in
this way we will be able to amend the existing results concerning
deformed strips.

First we introduce a unifying notation. Let~$n=2,3$ be the
dimension of the considered deformed region, \ie, the perturbed
planar strip or layer, respectively. We set~$x=(x^1,\dots,x^{n-1})
\in\Real^{n-1}$ and~$(x,u)\in\Omega_0:=\Real^{n-1}\times (0,d)$
for the unperturbed domain. From technical reasons it is
convenient to change the setting slightly, in comparison
with~(\ref{layer}) and~\cite{BGRS}, \cite{EV3}, and to deform the
``lower'' boundary of $\Omega_0$ what we certainly can do without
loss of generality. We denote therefore in this section
\begin{equation}\label{mirror}
  \Omega_\lambda:=\left\{(x,u)\in\Real^n :\,-\lambda d v(x)<u<d\right\}
\end{equation}
with~$v\in\Comp(\Real^{n-1})$. We denote by $-\Delta'$
the~$(n-1)$-dimensional Laplacian, while $-\Delta$ stands for
the~$n$-dimensional one. We also use
$$
  \langle f \rangle := \int_{\Real^{n-1}} f(x)\,dx\,,
$$
$\|\cdot\|$ as the norm in~$\sii(\Real^{n-1})$, and
$$
  \alpha(m):=\left\{
  \begin{array}{l}
    m \\
    (\ln m)^{-1}
  \end{array}
  \right.
  \qquad
  \beta(t):=\left\{
  \begin{array}{lll}
    t     & \qquad\textrm{if} & n=2 \\
    \ln t & \qquad\textrm{if} & n=3
  \end{array}
  \right.
$$

\subsection{The asymptotic expansion}

Let us now construct the asymptotics of the eigenvalues
$m_\lambda$ of the following boundary value problem:
\begin{eqnarray*}
&& (\Delta+\kappa_1^2)\Psi_\lambda=m_\lambda^2\Psi_\lambda
  \quad\textrm{in}\ \Omega_\lambda \\
&& \Psi_\lambda\left(x,\lambda d v(x)\right)=\Psi_\lambda(x,d)=0
\end{eqnarray*}
as they approach zero. We will seek it in the form
\begin{equation}\label{m-as}
  m_\lambda=\left\{
  \begin{array}{lll}
    \sum_{i=1}^\infty \lambda^i m_i
    & \textrm{if} & n=2 \\
    && \\
    \exp{\left(-\left(\sum_{i=1}^\infty \lambda^i m_i\right)^{-1}\right)}
    & \textrm{if} & n=3 \\
  \end{array}
  \right.
\end{equation}
where the existence of such expansions follows from \cite{BGRS}
and Theorem~\ref{thm.expansion}, respectively. Notice that this
corresponds to the expansion of~$E(\lambda) =\kappa_1^
2-m_\lambda^2$, the ground-state eigenvalue of
$-\Delta_D^{\Omega_\lambda}$ in the problem discussed above,
because the mirror transformation of $\Omega_\lambda$ on
(\ref{mirror}) does not affect the spectral properties.

Suppose that a function~$f\in\Comp(\Real^{n-1})$, $\supp f \cap
\supp v=\emptyset$, and $\langle f \rangle\not=0$ is given. If we
manage to construct a solution~$\psi_\lambda(x,u;m)$ of the
boundary value problem
\begin{eqnarray}\label{bv-basic}
  (\Delta+\kappa_1^2)\psi_\lambda=m^2\psi_\lambda
  +\left(\alpha(m)-\alpha(m_\lambda)\right) f \chi_1
&\textrm{in}& \Omega_\lambda \\
  \psi_\lambda=0
&\textrm{on}& \partial\Omega_\lambda \nonumber
\end{eqnarray}
which is bounded and non-vanishing w.r.t. $m$ for small nonzero $m$,
then~$\Psi_\lambda(x,u)=\psi_\lambda(x,u;m_\lambda)$. We shall look for
the asymptotics of~$\psi_\lambda$ in the following form,
\begin{equation}\label{psi-as}
  \psi_\lambda(x,u;m)=\sum_{i=0}^\infty \lambda^i \psi_i(x,u;m)\,.
\end{equation}
Substituting~(\ref{psi-as}) and~(\ref{m-as})
into~(\ref{bv-basic}), we obtain a family of the boundary value
problems:
\begin{eqnarray}
  (\Delta+\kappa_1^2)\psi_0=m^2\psi_0
  +\alpha(m) f \chi_1
&\textrm{in}& \Omega_0
  \qquad\quad i=0 \label{bv-chain0} \\
  \psi_0=0
&\textrm{on}& \partial\Omega_0 \nonumber \\
&& \nonumber \\
 (\Delta+\kappa_1^2)\psi_i=m^2\psi_i
  +(-1)^{n-1} m_i f \chi_1
&\textrm{in}& \Omega_0
  \qquad\quad i\geq 1 \label{bv-chaini} \\
  \psi_i=0
&\textrm{if}& u=d \nonumber \\
  \psi_i=-\sum_{j=1}^i \frac{d^j (-v)^j}{j!}
  \frac{\partial^j\psi_{i-j}}{\partial u^j}
&\textrm{if}& u=0 \nonumber
\end{eqnarray}
One can check easily that $\psi_0=-\alpha(m)(-\Delta' +m^2)^{-1} f
\chi_1$ solves~(\ref{bv-chain0}) and has the asymptotics
\begin{eqnarray} \label{as0}
\lefteqn{\psi_0(x,u;m)=
  \frac{(-1)^{n-1}}{2\pi^{n-2}}\,\chi_1(u)
  \Bigg[\langle f \rangle} \nonumber \\
&& +(-1)^{n-1}\alpha(m) \left(\int_{\Real^{n-1}} \beta(|x-x'|) f(x')\,dx'
  +\delta_n^3 (\gamma-\ln 2) \langle f \rangle\right) \nonumber \\
&& +\mathcal{O}\left(\alpha(m)^2 \right) \Bigg]
\end{eqnarray}
as $m\to 0$, where $\gamma$ is the Euler number
and~$\delta_n^j$ the Kronecker delta.
\begin{lemma}\label{lemma3.1}
Suppose that $F\in\Diff(\overline{\Omega_0})$ with a bounded support
and $H\in\Comp(\Real^{n-1})$ have the expansions
$$
  F(x,u;m)=\sum_{i=0}^\infty \alpha(m)^i F_i(x,u)\,,
  \qquad
  H(x;m)=\sum_{i=0}^\infty \alpha(m)^i H_i(x)
$$
as~$m\to 0$. Define
 $
  F_{i,k}:=\int_0^d F_i(\cdot,u)\,\chi_k(u)\,du\,.
 $
Let~$\phi_0$ be the solution of the boundary value problem
\begin{eqnarray}\label{bv-lemma0}
  (\Delta+\kappa_1^2)\phi_0=F_0
&\textrm{in}& \Omega_0\,, \\
  \phi_0=0
&\textrm{if}& u=d\,, \nonumber \\
  \phi_0=H_0
&\textrm{if}& u=0\,; \nonumber
\end{eqnarray}
then the condition
\begin{equation}\label{condition}
  \langle F_{0,1} \rangle
  =\sqrt{\frac{2}{d}}\,\kappa_1\,\langle H_0 \rangle
\end{equation}
is necessary and sufficient for existence of a solution of the
boundary value problem
\begin{eqnarray*}
  (\Delta+\kappa_1^2)\phi=m^2\phi+F
&\textrm{in}& \Omega_0\,, \\
  \phi=0
&\textrm{if}& u=d\,, \\
  \phi=H
&\textrm{if}& u=0\,,
\end{eqnarray*}
which is bounded as~$m\to 0$. If it is satisfied, the solution has
the asymptotics
\begin{eqnarray*}
  \lefteqn{ \phi(x,u;m) } \\ &&
  =\phi_0(x,u)+\frac{(-1)^{n-1}}{2\pi^{n-2}}
  \,\chi_1(u) \left(\langle F_{1,1} \rangle
  -\sqrt{\frac{2}{d}}\,\kappa_1\,\langle H_1 \rangle \right)
  +\mathcal{O}\left(\alpha(m)\right)\,.
\end{eqnarray*}
\end{lemma}
\PF The statement is obvious if ~$H=0$. In particular,
the solution~$\phi$ is constructed by the Fourier method
in the explicit form
$$
  \phi(x,u;m)=\sum_{i=1}^\infty \tilde{\phi}_i(x;m) \chi_i(u).
$$
By a direct calculation it is easy to see that~$\tilde{\phi}_i$
are bounded functions for~$m\geq 0$ so long as~$i\geq 2$. The
problem arises for~$i=1$, because in general  $\tilde{\phi}_1$
tends to infinity as~$m\to 0$. The condition~(\ref{condition})
guarantees that the explicit solution~$\phi$ has no such pole.
This proves the sufficiency. To see that the condition is
necessary at the same time, one integrates by parts in the
scalar product equation
$$
  \left(\chi_1, (\Delta+\kappa_1^2-m^2)\phi\right)
  =\left(\chi_1,F\right)
$$
and puts~$m=0$ afterwards. In the opposite case, $H\not=0$, we
use the replacement
$$
  \phi(x,u;m)=\varphi(x,u;m)
  +\left(1-\frac{u}{d}\right) H(x;m)
$$
and expand the r.h.s. of the equation for~$\varphi$ in the
Fourier series, which reduces the task to the previous
situation.\qed

\begin{corol}\label{Corol.of.Lemma3.1}
$\;\phi\in\Diff(\overline{Q})$ holds for any bounded
domain~$Q\subset\Omega_0$.
\end{corol}
\vspace{1mm}

\noindent It follows from Lemma~\ref{lemma3.1} that the
recursive system of the boundary value problem~(\ref{bv-chaini})
has solutions which are continuous with respect to $m$ in the
vicinity of $m=0$ and decay as~$|x|\to\infty$ for~$m>0$,
provided the~$m_i$'s satisfy the following recursive relations:
\begin{equation}
  m_i=(-1)^n \sqrt{\frac{2}{d}}\,\frac{\kappa_1}{\langle f \rangle}
  \sum_{j=1}^i \left\langle \frac{d^j (-v)^j}{j!}
  \frac{\partial^j\psi_{i-j}}{\partial u^j}(\cdot,0;0)
  \right\rangle.
\end{equation}
In particular, owing to~(\ref{as0}) and Lemma~\ref{lemma3.1} we
get
\begin{equation}\label{pole1}
  m_1=\frac{\kappa_1^2}{\pi^{n-2}}\, \langle v \rangle\,,
\end{equation}
which agrees with the leading term obtained by the
Birman-Schwinger method in the previous section --
\cf~Theorem~\ref{thm.expansion} and~(\ref{m-as}) -- as well as
with the corresponding result~(\ref{strip-expansion}) in the
strip case.

\subsection{The next-to-leading order}

Let us now calculate~$m_2$. By virtue
of~(\ref{bv-chaini}),~(\ref{as0}) and~(\ref{pole1}) the boundary
value problem for~$\psi_1$ together with the boundary condition
for~$\psi_2(x,u;0)$ look as follows
\begin{eqnarray}\label{eq1}
  (\Delta+\kappa_1^2)\psi_1=m^2\psi_1
  +(-1)^{n-1}\frac{\kappa_1^2}{\pi^{n-2}}\,\langle v \rangle\, f \chi_1
&\textrm{in}& \Omega_0 \\
  \psi_1=0
&\textrm{if}& u=d \nonumber \\
  \psi_1=d v \frac{\partial\psi_0}{\partial u}
&\textrm{if}& u=0 \nonumber \\
  \psi_2=0
&\textrm{if}& u=d \nonumber \\
  \psi_2=d v  \frac{\partial\psi_1}{\partial u}
&\textrm{if}& u=d,\ m=0 \nonumber
\end{eqnarray}
with
\begin{equation}\label{dpsi0}
\frac{\partial\psi_0}{\partial u}(x,0;m)=
  \frac{(-1)^{n-1}}{2\pi^{n-2}}\sqrt{\frac{2}{d}}\,\kappa_1
  \mathcal{B}(f) \,,
\end{equation}
where $\mathcal{B}(f)$ is the square bracket from (\ref{as0}).
Hence
\begin{equation}\label{pole-m2}
  m_2=(-1)^{n-1} \sqrt{\frac{2}{d}}\,\frac{\kappa_1 d}{\langle f \rangle}
  \left\langle v \frac{\partial\psi_1}{\partial u}(\cdot,0;0) \right\rangle
\end{equation}
and it is sufficient to find~$\psi_1$. With eq.~(\ref{eq1}) and
Lemma~\ref{lemma3.1} in mind, we consider the following boundary
value problem
\begin{eqnarray}\label{eq-phi0}
  (\Delta+\kappa_1^2)\phi_0=(-1)^{n-1}\frac{\kappa_1^2}{\pi^{n-2}}
  \,\langle v \rangle\, f \chi_1
&\textrm{in}& \Omega_0 \\
  \phi_0=0
&\textrm{if}& u=d \nonumber \\
  \phi_0=\frac{(-1)^{n-1}}{2\pi^{n-2}}\sqrt{\frac{2}{d}}\,\kappa_1
  \,d\,v\,\langle f \rangle
&\textrm{if}& u=0 \nonumber
\end{eqnarray}
and seek~$\phi_0$ in the form
\begin{equation}\label{phi0}
  \phi_0(x,u)=\frac{(-1)^{n-1}}{2\pi^{n-2}}\sqrt{\frac{2}{d}}\,\kappa_1
  \left[\left(1-\frac{u}{d}\right) \langle f \rangle\,d\,v(x)
  -\varphi(x,u)\right];
\end{equation}
substituting it into~(\ref{eq-phi0}), we arrive at the boundary
value problem
\begin{eqnarray*}
  (\Delta+\kappa_1^2)\varphi
  =-d\,\langle f \rangle \left(1-\frac{u}{d}\right)
  (\Delta'+\kappa_1^2)v+2\kappa_1\sqrt{\frac{2}{d}}
  \ \langle v \rangle\,f \chi_1
&\textrm{in}& \Omega_0 \\
  \varphi=0
&\textrm{on}& \partial\Omega_0.
\end{eqnarray*}
The Fourier method gives
\begin{eqnarray*}\label{varphi}
  \varphi
&=& -\sqrt{\frac{2}{d}}\,d\,\langle f \rangle
  \sum_{k=2}^\infty \frac{\chi_k}{\kappa_k}\,
  (-\Delta'+\kappa_k^2-\kappa_1^2)^{-1} (-\Delta'-\kappa_1^2)\,v \\
&& -\sqrt{\frac{2}{d}}\,d\,\frac{\chi_1}{\kappa_1}
  \left[\langle f \rangle\,v+\kappa_1^2 (-\Delta')^{-1}
  \left(\langle v \rangle\,f-\langle f \rangle\,v\right)\right].
\end{eqnarray*}
Lemma~\ref{lemma3.1} an relations (\ref{eq1}), (\ref{dpsi0}),
(\ref{eq-phi0}), and (\ref{phi0}) together with the last result
imply that
\begin{eqnarray*}
\lefteqn{\frac{\partial\psi_1}{\partial u}(x,0;0)
  =\frac{(-1)^n}{2\pi^{n-2}}\,\sqrt{\frac{2}{d}}\,\kappa_1} \\
&& \times \Bigg\{\frac{\kappa_1^2}{\pi^{n-2}} \Bigg[
  \ \int\limits_{\Real^{n-1}\times\Real^{n-1}} \!\!\!\!\!\!
  v(x)\,\beta(|x-x'|)\,f(x')\,dx\,dx' \\
&& +\langle f \rangle \int_{\Real^{n-1}} \beta(|x-x'|)\,v(x')\,dx'
  -\langle v \rangle \int_{\Real^{n-1}} \beta(|x-x'|)\,f(x')\,dx'
  \Bigg] \\
&& +\langle f \rangle \Bigg[3\,v(x)
  +2\sum_{k=2}^\infty \left[(-\Delta'+\kappa_k^2-\kappa_1^2)^{-1}
  (-\Delta'-\kappa_1^2)\,v\right](x) \\
&& +\delta_n^3\,\frac{\kappa_1^2}{\pi}\,(\gamma-\ln 2)
  \,\langle v \rangle \Bigg]\Bigg\}\,,
\end{eqnarray*}
where we have employed also the implication
$$
  \langle F \rangle=0 \ \Rightarrow\
  (-\Delta')^{-1}F=\frac{-1}{2\pi^{n-2}}
  \int_{\Real^{n-1}} \beta(|\cdot-x'|)\,F(x')\,dx'.
$$
Substituting this into~(\ref{pole-m2}) we get the sought
coefficient:
\begin{eqnarray}\label{m2}
m_2
&=& -\frac{\kappa_1^2}{\pi^{n-2}} \Bigg\{3\,\langle v^2 \rangle
  +\frac{\kappa_1^2}{\pi^{n-2}}
  \!\!\!\!\int\limits_{\Real^{n-1}\times\Real^{n-1}} \!\!\!\!\!\!
  v(x)\,\beta(|x-x'|)\,v(x')\,dx\,dx' \nonumber \\
&& +2\left\langle v \sum_{k=2}^\infty
  (-\Delta'+\kappa_k^2-\kappa_1^2)^{-1} (-\Delta'-\kappa_1^2)\,v
  \right\rangle \nonumber \\
&& +\delta_n^3\,\frac{\kappa_1^2}{\pi}\,(\gamma-\ln 2)
  \,\langle v \rangle^2 \Bigg\}.
\end{eqnarray}

\subsection{The critical case}

As we have pointed out in the introduction, the above result is most
interesting in the critical case, $\langle v \rangle=0$, when the first
coefficient~(\ref{pole1}) equals zero and~$m_2$ given by (\ref{m2})
determines the leading order. In this situation we have the following
result.
\begin{thm}
Let~$V\in\Comp(\Real^{n-1})$ be an arbitrary function
such that $\langle V \rangle=0$ and
$$
  v(x)=V\left(\frac{x}{\sigma}\right),
  \qquad \sigma>0\,.
$$
Then the following inequalities hold,
\begin{multline*}
  -\frac{\kappa_1^2\sigma^{n-1}}{\pi^{n-2}}
  \left(\frac{8}{2}\|V\|^2
  +\frac{3}{2\kappa_1^2\sigma^2} \|V\| \|\Delta' V\|
  -2\kappa_1^2\sigma^2 \|\nabla'(\Delta')^{-1}V\|^2\right)
  \\
\leq m_2 \leq -\frac{\kappa_1^2\sigma^{n-1}}{\pi^{n-2}}
  \left(\frac{3}{2} \|V\|^2
  -2\kappa_1^2\sigma^2 \|\nabla'(\Delta')^{-1}V\|^2\right)\,.
\end{multline*}
\end{thm}
\PF In the first place, note that $\langle V \rangle=0$ implies
$$
  -\frac{\kappa_1^2}{\pi^{n-2}}
  \!\!\!\!\int\limits_{\Real^{n-1}\times\Real^{n-1}} \!\!\!\!\!\!
  V(x)\,\beta(|x-x'|)\,V(x')\,dx\,dx'
  =\|\nabla'(\Delta')^{-1}V\|^2>0\,,
$$
because~$\Delta'\beta(|x|)=2\pi^{n-2}\delta(x)$ holds in the sense
of distribution. Under the stated assumptions, the
formula~(\ref{m2}) yields therefore
$$
  m_2=-\frac{\kappa_1^2\sigma^{n-1}}{\pi^{n-2}}
  \bigg(3 \|V\|^2-2\kappa_1^2\sigma^2 \|\nabla'(\Delta')^{-1}V\|^2
  +2 A(\sigma)\bigg),
$$
where
$$
  A(\sigma):=\sum_{k=2}^\infty \left\langle V
  \left(-\Delta'+(\kappa_k^2-\kappa_1^2)\,\sigma^2\right)^{-1}
  (-\Delta'-\kappa_1^2\sigma^2)\,V
  \right\rangle\,,
$$
and it suffices to find suitable bounds on~$A(\sigma)$.

Since the Fourier transformation together with the Plancherel
theorem give the estimate
\begin{equation}\label{Fourierbound}
  \left\|\left(-\Delta'+(\kappa_k^2-\kappa_1^2)\,
  \sigma^2\right)^{-1}F\right\|
  \leq \frac{\|F\|}{(\kappa_k^2-\kappa_1^2)\sigma^2}\,,
\end{equation}
we obtain the upper bound
$$
  A(\sigma) \leq \frac{3}{4}\left(\|V\|^2
  +\frac{1}{\kappa_1^2\sigma^2} \|V\| \|\Delta'V\|
  \right),
$$
where the numerical factor comes
from~$\sum_{k=2}^\infty(k^2-1)^{-1}=\frac{3}{4}$.

On the other hand, denoting
$$
  U_k(x;\sigma):=\left[\left(-\Delta'
  +(\kappa_k^2-\kappa_1^2)\,\sigma^2\right)^{-1}V\right](x),
$$
we see that
\begin{eqnarray*}\label{Uk}
\lefteqn{ \left\langle V
  \left(-\Delta'+(\kappa_k^2-\kappa_1^2)\,\sigma^2\right)^{-1}
  (-\Delta'-\kappa_1^2\sigma^2)\,V \right\rangle } \\ &&
=\Big\langle \left(-\Delta'+(\kappa_k^2-\kappa_1^2)\,\sigma^2\right)
  U_k (-\Delta'-\kappa_1^2\sigma^2) U_k \Big\rangle\,.
\end{eqnarray*}
Integrating the \rhs\/ by parts and using~(\ref{Fourierbound}), we
get the lower bound
{\setlength\arraycolsep{2pt}
\begin{eqnarray*}
  A(\sigma)
&=& \sum_{k=2}^\infty \Big(\|\Delta' U_k\|^2
  +\kappa_1^2(k^2-2)\sigma^2 \|\nabla'U_k\|^2
  -\kappa_1^4(k^2-1)\sigma^4 \|U_k\|^2\Big) \\
&>& -\sum_{k=2}^\infty \kappa_1^4(k^2-1)\sigma^4 \|U_k\|^2
  \geq - \|V\|^2 \sum_{k=2}^\infty \frac{1}{k^2-1}
  =-\frac{3}{4} \|V\|^2,
\end{eqnarray*}}\noindent
which concludes the proof.\qed \vspace{2mm}

\noindent This theorem confirms the spectral picture we got from
(\ref{crit-nonexist}) and (\ref{crit-exist}). More specifically, $m_2>0$
as $\sigma\to\infty$ so the critical weakly bound state exists for
sufficiently smeared deformations, and vice versa. In contrast to
(\ref{crit-nonexist}) and (\ref{crit-exist}), however, we are able now to
tell from (\ref{m2}) for any given zero-mean $v$ the sign of $m_2$.

\subsection*{Acknowledgment}

R.G. is grateful for the hospitality extended to him at NPI AS
where a part of this work was done. The research has been
partially supported by GA AS and the Czech Ministry of Education
under the contracts 1048801 and ME170. The first and the third
authors have been partially supported by Russian Fund of Basic
Research -- Grants 99-01-00139 and 99-01-01143, respectively.


\providecommand{\bysame}{\leavevmode\hbox to3em{\hrulefill}\thinspace}

\vspace{10mm}

\begin{flushleft}
R. Borisov and R. Gadyl'shin \\ Bashkir State Pedagogical
University \\ October Revolution St.~3a\\ RU-450000 Ufa, Russia
\\ Email: borisovDI@ic.bashedu.ru,
%
%
gadylshin@bspu.ru \\[5mm]

P. Exner and D. Krej\v{c}i\v{r}\'{\i}k \\ Department of
Theoretical Physics \\ Nuclear Physics Institute \\ Academy of
Sciences \\ CZ-25068 \v Re\v z, Czech Republic \\ Email:
exner@ujf.cas.cz, krejcirik@ujf.cas.cz

\end{flushleft}

\end{document}